\documentclass[12pt]{article}   
 \usepackage{psfig,times,fancyheadings}
        
\begin{document}
\thispagestyle{empty} 


 \lhead[\fancyplain{}{\sl }]{\fancyplain{}{\sl }}
 \rhead[\fancyplain{}{\sl }]{\fancyplain{}{\sl }}


\newcommand{\nc}{\newcommand}

\nc{\qI}[1]{\section{ {#1} }}
\nc{\qA}[1]{\subsection{ {#1} }}
\nc{\qun}[1]{\subsubsection{ {#1} }}
\nc{\qa}[1]{\paragraph{ {#1} }}

\nc{\qfoot}[1]{\footnote{ {#1} }}
\def\qL{\hfill \break}
\def\qpar{\vskip 2mm plus 0.2mm minus 0.2mm}

\def\qparr{ \vskip 1.0mm plus 0.2mm minus 0.2mm \hangindent=10mm
\hangafter=1}

\def\qdec#1{\par {\leftskip=2cm {#1} \par}}

\def\qdpt{\partial_t}
\def\qdpx{\partial_x}
\def\qddpt{\partial^{2}_{t^2}}
\def\qddpx{\partial^{2}_{x^2}}
\def\qn#1{\eqno \hbox{(#1)}}
\def\qds{\displaystyle}
\def\qal{\sqrt{1+\alpha ^2}}
\def\qw{\widetilde}


\null
\vskip 3cm

\centerline{\bf \LARGE Identifying the bottom line after a stock market crash}
\vskip 0.5cm
 \centerline{\bf \Large }

\vskip 1cm
\centerline{\bf B.M. Roehner $ ^* $ }
\centerline{\bf L.P.T.H.E. \quad University Paris 7 }

\vskip 2cm

{\bf Abstract}\ In this empirical paper we show that in the months following
a crash there is a distinct connection between the fall
of stock prices and the increase in the range of interest rates for a sample
of bonds. This variable, which is often referred to as the interest rate
spread variable, can be considered as a statistical measure for the disparity
in lenders' opinions about the future; in other words, it provides an operational
definition of the uncertainty faced by economic agents. The observation that
there is a strong negative correlation between stock prices and the spread 
variable relies on the examination of 8 major crashes in the
United States between 1857 and 1987.  That relationship which has remained 
valid for one and a half century in spite of important changes in the 
organization of financial markets can be of interest in the perspective
of Monte Carlo simulations of stock markets.

\vskip 1cm
  \centerline{\bf  23 September  1999 }

\vskip 1cm
{\bf Key words}\quad Stock market - Crashes - Uncertainty - 
 Interest rate spread - Comparative study

\vskip 2cm
* Postal address: LPTHE, University Paris 7, 2 place Jussieu, 75005 Paris, France
\qL
\phantom{* }E-mail: ROEHNER@LPTHE.JUSSIEU.FR
\qL
\phantom{* }FAX: 33 1 44 27 79 90

\vfill \eject

\qI{Stock prices and interest rates}

In this paper we show that in the time interval between crash and recovery
there
is a clear relationship between price variations and the dispersion of 
interest rates for bonds of different grades (see below), 
i.e. what is usually called the
interest rate spread. Before explaining this relationship in more detail
let us emphasize that it was observed empirically from the 
mid-nineteenth century 
to the latest major crash in 1987. This is in strong contrast with so
many ``regularities'' which are dependent upon specific business circumstances.
Such is for instance the case of the interest rate 
itself.\qL
Because of the close connection between stock
and bond markets one would 
expect a strong link between stock prices and interest rates. This is not the case however; there seems
to be no {\it permanent} relationship between these variables; see in this 
respect the conclusions of [13] and [18, p.241]. 
It is true that sometimes a slight decrease in interest rates,
by changing the ``mood'' of the market, suffices to send prices upward. 
Thus, in the
fall of 1998 three successive quarter point decreases of the federal-fund
rate (that is to say a global -0.75\%)
stopped the fall of the prices and brought about a rally. 
In other circumstances, however, even a huge drop in interest rates is
unable to stop the fall of stock prices; an example is provided by the period
from January 1930 to May 1931 when the interest rate fell from 6\% to 2\%
without any effect on the level of stock prices; similarly in the
aftermath of the 1990 crash of the Japanese stock market interest rates went down
to almost zero percent without bringing about any recovery. \qL
One should not be surprised by the changing relationship between stock price
levels and interest rates. Something similar can be observed in meteorology:
sometimes a small fall in temperature is sufficient to produce rain, while
in other circumstances a huge fall in  temperature will not give any rain. 
In this case we know that the phenomenon has something to do with the
hygrometric degree of the air; in the case of the stock market we do not
really know which one of the many other variables plays the 
crucial role. 
In the light of such changing patterns the fact that the relationship between
stock prices and the spread variable appears to be so robust and so stable in 
the course of time is worthy of attention. 

\qI{Interest rate spread and uncertainty}
It is a common saying that ``markets dislike and fear uncertainty''. In a 
strong bull market there is little uncertainty; for everybody the word of the
day is ``full steam ahead''. The situation is completely different after
a crash. There is uncertainty about the duration of the bear market; some 
would think that it will be short while others expect a long crisis.
In 1990 when the bubble burst on the Tokyo stock market only few people
would probably have expected the crisis to last for almost ten years. 
There is also uncertainty about which sectors will be the first to emerge
from the turbulence: banks or investment funds, property funds or technology
industry, etc. \qL
As we know the interest rate represents the price a company pays to buy 
money for the future. The more uncertain the future, the riskier the
investment, the higher the interest rate. We will indeed see that during
recessions interest rates often (but not always)
show an upward trend. In addition, and this
is probably even more important, the increased uncertainty produces 
greater disparity in the rate of different loans. This uncertainty has 
different sources 
(i)those who expect a short
crisis will be tempted to lend at lower rates than those who fear a protracted
recession 
(ii) the fact that there is no longer any ``leading force'' in 
the economy obscures expectations; therefore it becomes more difficult to 
make a reliable risk assessment 
for low-quality borrowers (representing the so-called low-grade
bonds). \qL
In short, the interest rate spread gives
us a means to probe the mood, expectations and forecasts of managers, a means
which is probably more reliable than the standard confidence indexes 
obtained from surveys (in this respect see the last section). \qL
Although in many
econophysical models of the stock market 
[3, 5, 7, 9, 10, 16] 
interest rates do not play a role per se, the fact
that uncertainty is greater in the downward phase of the speculative cycle
than in the upward phase could be built into the models by adjusting the
randomness of the stochastic variables used in Monte Carlo simulations. 
In contrast, interest rates usually play a determinant role in 
econometric models. A particularly attractive model of that kind is the
Levy-Levy-Solomon model;  
it describes the stock and bond markets as communicating vessels and how
traders switch from one to the other. 
The book by Oliveira et al.
[12, chapter 4] details  
the assumptions of the model,
and, through simulations, explains how it works and to which results it leads.

\qI{The data}
Monthly stock price data going back into the 19th century 
can be found
fairly easily; possible sources are [4, 8, 17]. Measuring
interest rate spreads is a more difficult matter. To begin with it is not
obvious which estimates should be used. The primary source about
bond rates is [8]; furthermore a procedure for constructing the
spread measure was proposed in [11]. As a matter of fact
Mishkin's stimulating paper provided the main incentive for the writing of
the present paper. Mishkin proposed to represent the spread by the difference
between the one-fourth of the bonds of the lowest grade (i.e. high rates)
and the one-fourth of the bonds of the best grade (i.e. low rates). 
It turns out that even for the mid-nineteenth century Macaulay's data provided
at least three bonds in each of these classes which is fairly sufficient to
give acceptable accuracy; for the more recent period
1888-1935 there are as many as 10 bonds in each ``quartile''. For post-World 
War II crashes, Macaulay's series can be prolonged by the data in [1]. 
More detailed comments about how these two measures compare can be found in
[11] 

\qI{Results}
\qA{Connection between share prices and interest rate spread between crash and
recovery}
Fig.1a and 1b show the evolution of stock prices (thick solid line), 
interest rate
spread (thick dashed line), and interest rate (thin dashed line) for 8 major 
crashes. The left-hand vertical scale is the same for all graphs except
1929: this allows a visual comparison of the crashes' severity. The right-hand
vertical scales although not identical (which was not possible due to 
different orders of magnitude) are nevertheless comparable in the sense 
that the overall range $ y_{\hbox{max}}/y_{\hbox{min}} $ is the same (except 
again for 1929); this allows a visual comparison of the increase of the 
spread. The horizontal scales represent  the number of months after the 
crash; these scales are the same for all graphs (with the exception of 1929);
this allows a comparison of the time elapsed between crash and 
recovery. \qL
It can be seen that the decline in stock prices is mirrored in a similar 
increase in interest rate spread. As a matter of fact the chronological
coincidence between the troughs of the stock prices and the peaks of the
spread variable is astonishing. Even for the 1929-1932 episode for
which there is a 30-month span between crash and recovery
the peak for the spread variable coincides almost to the month with the end
of the price fall. \qL
The connection between both variables is confirmed by the correlation 
coefficients (left-hand correlations in Fig.1): they are all negative and
comprised between $ -0.64 $ and $ -0.94 $; note that the smallest correlation
($ -0.64 $) corresponds to a relatively small crash with a fall in 
stock prices of less than 20\%. \qL
For 19th century episodes the interest rate changes are more or less in the
same direction as those of the spread variable; however the correlations with
stock prices (right-hand correlations in Fig.1) are substantially 
lower. For 20th century episodes the picture changes completely: the
interest rate no longer moves in the same direction as the spread variable; 
consequently these correlations become completely random in contrast to the 
correlations between stock prices and spread variable which remain close to 
$ -1 $. 
In the interpretative framework that we developed above we come up with the 
following picture. After a crash uncertainty, doubts and apprehension begin to
spread throughout the market; usually (leaving 1929 apart for the moment) the 
fall last about 10 months; during that time, uncertainty continues to 
increase. Then, suddenly, within one month, the trend shifts in the opposite
direction: price begin to increase and uncertainty to subside. \qL
One may wonder how the spread variable behaved in the bull phases. First of
all one should note that not all the crashes that we examined were preceded
by a wild bull market; so we concentrate here on  
three typical bull markets that occurred in 1904-1907,
1921-1929 and 1985-1987. During these periods the spread variable remained
almost unchanged. Similarly during the period 1950-1967 which was marked 
by a considerable increase in stock prices (without however being followed 
by a major crash) the spread variable remained at a fairly constant level 
of 1.5\%. In contrast during the period 1968-1979 which was marked by a 
downward trend in stock prices the spread variable was substantially 
larger in the range 2.5\% -3.8\%. \qL
A simple look at the charts in Fig.1 confirms what we already know,
namely that the crisis of 1929-1932 was quite exceptional. This is
of course obvious in economic terms (unemployment, drop in industrial
production, etc.); it is also true from a purely financial perspective.
Stock prices plummeted from a level 100 to less than 20, and the spread
variable increased from 2.5\% to almost 8\%, a three-fold increase. For other
episodes (see table 1) the corresponding ratios are all below 1.85. As an
illustration of the intensity of the financial crisis one can mention the fact
that November and December 1929 saw the failure of  $ 608 $ banks; the 
crisis continued in subsequent months to the extent that in March 1933 one
third of all American banks had disappeared ([11]).  

\vskip 1cm 

\centerline {\bf Table 1 \quad Stock price changes versus 
increase in interest rate spread}

$$ \matrix{
\hbox{Year}     & \hbox{Stock price}  & \hbox{Interest rate spread} \cr
\hbox{of crash} & \hbox{fall}       & \hbox{increase} \cr
   & A_{\hbox{price}} &  A_{\hbox{spread}} \cr
   &  &   \cr
1857 & 1.63 & 1.46 \cr
1873 & 1.24 & 1.32 \cr
1890 & 1.23 & 1.09 \cr
1893 & 1.34 & 1.38 \cr
1906 & 1.46 & 1.82 \cr
1929 & 6.12 & 3.05 \cr
1937 & 1.89 & 1.82 \cr
1987 & 1.40 & 1.25 \cr
} $$

\vskip 1cm
where:
$$  A_{\hbox{price}}= \hbox{peak price / minimum price}, \quad
A_{\hbox{spread}}= \hbox{maximum spread / initial spread} $$

If we leave 1929 apart the fall/increase ratios of the two variables are
almost of the same magnitude; a linear fit gives:
$$ A_{\hbox{price}} = \alpha A_{\hbox{spread}} + \beta $$

with: $ \alpha =0.63 \pm 0.50, \beta = 0.54 \pm 0.13 $, the correlation
is equal to $ r=0.74 $ 
(confidence interval for $ r $ to probability 0.95 is 0. to 0.96). \qL
If we include 1929 in the sample the coefficients of the linear fit
change completely and become:
$ \alpha =2.53 \pm 0.67, \beta = -2.11 \pm 0.39 $, with a correlation equal
to $ 0.96 $ (confidence interval to probability 0.95 is 0.74 to 0.99). Needless
to say the last fit has to be looked upon with cautiousness since it is so much
dependent upon the figures of the 1929 crash. 

\qA{Connection between interest spread and market's uncertainty}

In section 3 we interpreted the spread variable as characterizing the 
uncertainty and lack of confidence existing in the market at a given moment. 
This interpretation was based on plausible arguments but one would be on
firmer ground if it could be supported by some statistical evidence. In this
paragraph we provide at least partial proof in that respect by comparing the
changes of the spread variable to the consumers' lack of confidence as measured
by standard surveys. This is shown in Fig. 2.; it represents the spread
variable along with the lack of confidence index in the 
United States in the period before and after the 1987 crash. Changes
in the two variables are fairly parallel although the spread variable appears
to be much more sensitive and displays larger fluctuations. In the two
months before the crash of 19 October 1987 both the uncertainty (measured by the
spread variable) and the lack of confidence (estimated through consumer
surveys) increased by about 20\%; after the crash both variables 
increased rapidly; but the after-effects of the crash
were short-lived and uncertainty
decreased after the beginning of 1988. If consumer confidence data
could be found for the period prior to World War II it would of course be 
interesting to perform a similar comparison for other crashes.

\qI{Perspectives for an extension to other speculative markets}
Relationships which have a validity extending over one century are not 
frequent either in economics or in finance. Yet, if the above observation 
remains isolated it will be hardly more than a technical feature
of interest for stock market professionals. 
It is tempting to
posit that an increase in uncertainty can play a similar role in 
other speculative markets. Stock markets are certainly special in so far
as they are {\it pure} speculative markets; in contrast to property or 
commodities, stocks do not have any other usage for their buyer
than to earn dividends. 
Nevertheless the stock market seems to be in close connection with the
property market; historically stock market crashes have often been 
preceded by a collapse of property prices; see in this respect 
[6, p.65]
and [14, p.76]. One problem with the property market is its
long relaxation time. For that reason we consider here another case
namely the market for gold, silver and diamonds. As is well known,
starting in 1977  huge speculative bubbles developed in these 
items, which collapsed simultaneously in January 1980. 
Let us concentrate on the diamond market since the gold market has
already been closely investigated 
particularly by A. Johansen and D. Sornette. In Fig.3 we represented
the price of diamonds along with the consumer lack of confidence index that
we already used above. Two observations can be made (i) There is a huge
increase in the lack of confidence index between 1978 and the spring of
1980 that is to say during the period when the bubble developed. This 
shows that it would be vain to explore the diamond market (or silver/gold
markets) in order to find specific causes for the collapse. It 
was most certainly triggered by exogenous, psycho-sociological factors. 
(ii) In the phase between collapse and recovery (March 1980-March 1986), in
contrast to what we observed with stock prices, there is no connection
whatsoever between diamond price changes and the fluctuations of the lack
of confidence index. Perhaps the story would be different if one could use
a confidence index specially pertaining to the diamond market.


\vfill \eject



\centerline{\bf \Large References}

\vskip 1cm

\qparr
(1) BERNANKE (B.S.) 1983: Non-monetary effects of the financial crisis in the
propagation of the Great Depression. American Economic Review 73,257.

\qparr
(2) BOUCHAUD (J.-P.), POTTERS (M.) 1997: Th\'eorie des risques financiers.
Alea-Saclay, Eyrolles. Paris.

\qparr
(3) CALDARELLI (G.), MARSILI (M.), ZHANG (Y.-C.) 1997: A prototype model of
stock exchange. Europhysics Letters 40,479.

\qparr
(4) FARREL (M.L.) 1972: The Dow Jones averages 1885-1970. Dow Jones. Princeton.

\qparr
(5) FEIGENBAUM (J.A.), FREUND (P.G.O.) 1996: Discrete scaling in stock markets
before crashes. International Journal of Modern Physics B 10,3737.

\qparr
(6) HARRISON (F.) 1983: The power in the land. An inquiry into unemployment,
the profits crisis and land speculation. Shepheard-Walwyn. London. 

\qparr
(7) LUX (T.), MARCHESI (M.) 1999: Scaling and criticality in a stochastic 
multi-agent model of a financial market. Nature 397, 498.

\qparr
(8) MACAULAY (F.R.) 1938: The movements of interest rates, bond yields and stock
prices in the United States  since 1856. National Bureau of Economic Research. 
New York. 

\qparr
(9) MANTEGNA (R.N.), STANLEY (H.E.) 1997: Stock market dynamics and turbulence:
parallel analysis of fluctuation phenomena. Physica A 239,255.

\qparr
(10) MANTEGNA (R.N.), STANLEY (H.E.) 1999: Scaling approach in finance. 
Cambridge University Press. Cambridge (in press). 

\qparr
(11) MISHKIN (F.S.) 1991: Asymmetric information and financial crises: a historical
perspective. in: Hubbard (R.G.): Financial markets and financial crises. 
National Bureau of Economic Research. University of Chicago Press. Chicago. 

\qparr
(12) OLIVEIRA (S.M. de), OLIVEIRA (P.M.C. de), STAUFFER (D.) 1999: Evolution,
money, wars and computers. Teubner. Stuttgart. See especially chapter 4
about stock market models.

\qparr
(13) OWENS (R.N.), HARDY (C.O.) 1929: Interest rates and stock speculation. A
study of the influence of the money market on the stock market. George Allen
and Unwin. London. 

\qparr
(14) ROEHNER (B.M.) 1999: Spatial analysis of real estate price bubbles: Paris,
1984-1993. Regional Science and Urban Economics 29,73.

\qparr
(15) SORNETTE (D.), JOHANSEN (A.) 1997: Large financial crashes. Physica A 245,411.

\qparr
(16) STAUFFER (D.), OLIVEIRA (P.M.C.), BERNARDES (A.T.) 1999: Monte Carlo simulation
of volatility correlation in microscopic market model. International
Journal for Theoretical and Applied Finance 2,83.

\qparr
(17) WILSON (J.), SYLLA (R.), JONES (C.P.) 1990: Financial market volatility. Panics
under the national banking system before 1914. Volatility in the long-run 
1830-1988. in White (E.N.) Crashes and panics: the lessons from history. 
Dow Jones-Irwin. Homewood.

\qparr
(18) WYCKOFF (P.) 1972: Wall Street and the stock markets. A chronology (1644-1971).
Chillon Book Company. Philadelphia. 

\vfill \eject

Figure captions
\qpar

{\bf Fig.1a \quad Stock prices versus interest rate spread: 19th century 
crashes.} Thick solid line: 
stock price index on the NYSE normalized to 100 at its peak value (left-hand 
vertical scale); thick dashed line: interest rate spread (right-hand vertical
scale). The thin dashed line represents the interest rate  for high grade 
commercial paper; it serves as a control variable in order to determine whether
it is the spread or the interest rate which is the pivotal variable. 
For the purpose of facilitating comparison
the left-hand vertical scale is the same for all graphs: this allows a visual comparison of the crashes' severity. The right-hand
vertical scales although not identical are nevertheless comparable in the sense 
that their overall ranges $ y_{\hbox{max}}/y_{\hbox{min}} $ are the same. The horizontal scales represent  the number of months after the 
crash; these scales are the same for all graphs. 
The numbers under the title are the correlations price/spread and price/interest 
rate respectively. {\it Sources: see text}. 

{\bf Fig.1b \quad Stock prices versus interest rate spread: 20th century 
crashes.} The caption is the same as for Fig.1a; note however that for the
1929 chart the
scales for the stock prices (right-hand vertical scale), for the spread 
(left-hand vertical scale) and for time (horizontal scale) are 
not the same as for the other charts. This
clearly shows the exceptional magnitude of the crash of 1929.
{\it Sources: see text}. 

{\bf Fig.2 \quad Comparison between the spread variable and the 
consumer lack of confidence index before the crash of October 1987.} 
Changes in the spread variable 
(solid line) and in the lack of confidence index (broken line) are fairly 
parallel but the first variable is much more sensitive. The lack of confidence
index is the inverse of the standard confidence index obtained from surveys. 
{\it Sources: Mishkin (1991), Gems and Gemology 24,140 (Fall 1998)}.

{\bf Fig.3 \quad Comparison of the price of diamonds before the collapse of
January 1980 with the evolution of the lack of confidence index}. In the
months before the market collapse the lack of confidence increased rapidly. 
However after the crash the lack of  confidence index does not show the same
pattern that we observed in Fig.1. The outcome would perhaps be different
if we could use a confidence index focused on the diamond market.
{\it Sources: Gems and Gemology 24 (Fall 1998)}.

\end{document}